\documentstyle[aps,epsf,float]{revtex}

\begin{document}
\draft

\title{
\hfill{\small{FZJ-IKP(TH)-1999-35}}\\[0.6cm]
On neutral pion electroproduction off
  deuterium\footnote{Work supported in part by Deutsche
  Forschungsgemeinschaft under contract no. Me864-16/1.}
}

\author{V.~Bernard,$^1$\footnote{E-mail:~bernard@lpt6.u-strasbg.fr}
H.~Krebs,$^2$\footnote{E-mail:~h.krebs@fz-juelich.de}
Ulf-G.~Mei{\ss}ner$^2$\footnote{E-mail:~Ulf-G.Meissner@fz-juelich.de}
}

\address{
$^{1}$Laboratoire de Physique Th\'eorique, 
Universit\'e Louis Pasteur, F-67084 Strasbourg Cedex, France \\
$^{2}$Institut f\"ur Kernphysik (Theorie), Forschungszentrum J\"ulich 
D-52425 J\"ulich, Germany 
}
\maketitle

\begin{abstract}
Threshold neutral pion electroproduction on the deuteron is studied in
the framework of baryon chiral perturbation theory at next--to--leading 
order in the chiral expansion. To this order in small momenta, the 
amplitude is finite and a sum of two-- and three--body interactions with no
undetermined parameters. We calculate the S--wave multipoles for
threshold production  and the deuteron S--wave cross section
as a function of the photon virtuality. We also discuss the
sensitivity to the elementary neutron amplitudes.
\end{abstract}


\medskip


\vspace{1cm}

\noindent Chiral perturbation theory has been successfully applied to
neutral pion photo-- and electroproduction off the 
proton~\cite{bkmz,bkme} as well as to $\pi^0$
photoproduction on the deuteron~\cite{bblmk}. The scattering off
deuterium is not only interesting {\it per} {\it se}, but also because
this loosely bound two--nucleon system can be used as a neutron
target. In particular, in ref.\cite{bblmk} it was shown that one can
indeed extract the elementary $\pi^0 n$ production amplitude from a
precise measurement on the deuteron. This was vindicated by the
experiment performed at SAL~\cite{SAL}. Furthermore, at MAMI experiments
for neutral pion electroproduction off deuterium at small photon virtualities
have been untertaken and are presently being analyzed~\cite{merkel}.
In this brief note we wish to report on
first results for this process obtained in chiral perturbation theory
to third order. We use the methodology developed by Weinberg~\cite{wein1},
which relates scattering processes involving a single
nucleon to nuclear scattering processes. The non-perturbative effects 
responsible for nuclear binding are accounted
for using phenomenological nuclear wavefunctions.\footnote{We are well
aware of recent developments in chiral effective field theories which
also provide fairly precise deuteron wavefunctions~\cite{ksw,egm}. However, for our
purpose the hybrid approach is suitable and we will comment on this
issue later on.} Although this
clearly introduces an inevitable model dependence, one can compute
matrix elements using a variety of wavefunctions in order to ascertain
the theoretical error induced by the off-shell behavior of different
wavefunctions. Here, we work to third order in the chiral expansion
and consider only threshold kinematics (i.e. the pion is produced at
rest) and thus calculate the pertinent transverse and longitudinal
S--wave multipoles. While a third order computation is not sufficient 
for the normalization of the elementary
amplitudes, the explicit calculations  for $\pi^0$ electroproduction
off the proton  to fourth
order let one expect that the momentum dependence of the S--wave
cross section is sufficiently accurately described at the order we are
working. This topic will be discussed in more detail below.

\medskip
\noindent To third order ($O({q^3})$, where $q$ denotes a
small momentum or a pion mass) in chiral perturbation
theory, the S--wave neutral pion electroproduction amplitude 
off the deuteron can be decomposed as follows:
\begin{equation}
{\cal M}_d = {\cal M}^{ss}_d + {\cal M}^{tb}_d =
 2\, {\cal M}_1 \, \vec{J}\cdot \vec{\epsilon} + 
2i\, {\cal M}_2 \,  \vec{J}\cdot\hat{k}\,\vec{J}\cdot ( \vec{\epsilon} 
\times \hat{k}) + 2\, {\cal M}_3 \,
 \vec{J}\cdot \hat{k} \,  \vec{\epsilon} \cdot \hat{k} ~,
\end{equation}
where $ \vec{J}$ is the deuteron angular momentum vector, and
$\vec{\epsilon}$ and $ \vec{k}$ are the polarization vector and
three--momentum of the virtual photon, respectively. Note that
in electron scattering $k^2 \le 0$. There are three multipole
amplitudes at threshold, which are of the electric, magnetic and
longitudinal type. These can be mapped onto the notation of ref.\cite{aren}
via ${\cal M}_1 = E_{01}^1 + M_{01}^2$, ${\cal M}_2 = 2M_{01}^2$ and
${\cal M}_1 + {\cal M}_3 = L_{01}^1$. In that notation, the upper
index gives the multipolarity, whereas the lower indices denote the
orbital angular and the total angular momentum of the final
deuteron--pion system, respectively. Effectively, however, one has
only two combinations of these multipoles contributing at threshold,
which only appear squared and are given by
\begin{equation}
|E_d|^2 \equiv |E_{01}^1|^2 + |M_{01}^2|^2~, \quad  |L_d|^2 \equiv |L_{01}^1|^2~.
\end{equation}
The electric dipole amplitude
$E_d$ characterizes the transverse response whereas $L_d$ parameterizes
the longitudinal response of the deuteron to the virtual photon. In general, the
multipoles depend on the photon virtuality $k^2$ and the pion energy
$\omega$. Since we only consider the production threshold $\omega_{\rm
  thr} = M_{\pi^0}$, we will not further specify this energy dependence.
In analogy to what is done in the single nucleon sector, we also consider
the deuteron S--wave cross section $a_{0d}$,
\begin{equation}
a_{0d} = |E_{d}|^2 + \epsilon_L \,  |L_{d}|^2~,
\end{equation}
with $\epsilon_L = -(k^2/k_0^2)\epsilon$ the longitudinal
polarization for a photon with polarization $\epsilon$.

\medskip\noindent
As shown in fig.~1, the amplitude obtains contributions from 
single scattering ($ss$) as well as the so--called three--body ($tb$)
graphs. The single scattering contribution for the transverse
and longitudinal multipoles at threshold takes the form
\begin{eqnarray}
E_d^{ss} &=& \frac{1+M_{\pi^0}/m_N}{1+M_{\pi^0}/m_d} \frac{1}{2} \, E_{0+} \,
S_d^1 (k^2)~, \nonumber\\
L_d^{ss} &=& \frac{1+M_{\pi^0}/m_N}{1+M_{\pi^0}/m_d} \frac{1}{2} \, \biggl\{
E_{0+} \bigl[S_d^1 (k^2) + S_d^2 (k^2) - S_d^3 (k^2) \bigr] +
L_{0+} \, S_d^3  (k^2) \,\biggr\}~,
\end{eqnarray}
where $M_\pi$, $m_N$ and $m_d$ denote the pion, nucleon and deuteron
mass, in order.\footnote{Note that we work with $M_{\pi^0} =
  134.97$~MeV and $M_{\pi^+} = 140.11$~MeV, to account for the
  neutron--proton mass difference in the rescattering diagrams. A
  detailed discussion of this point id given  in ref.\cite{bkmz}.}
Furthermore, 
\begin{equation}
X_{0+} = X_{0+}^{\pi^0 p} + X_{0+}^{\pi^0 n}~, \quad X = \{E,L\}~, 
\end{equation}
and the $S_d^i (k^2)$ are the pertinent deuteron form factors
\begin{eqnarray}
\int d^3 p \, \phi^* (\vec{p}\,) \,  \vec{S}\cdot \vec{\epsilon} \,
\phi (\vec{p} -\vec{k}/2) &=& S_d^1 (k^2)\,  \vec{J}\cdot \vec{\epsilon} 
+ S_d^2 (k^2) \,  \vec{J}\cdot \hat{k} \,  \vec{\epsilon} \cdot
\hat{k}~, \nonumber \\
\int d^3 p \, \phi^* (\vec{p}\,) \, \vec{S} \cdot \hat{k} \,
\vec{\epsilon} \cdot \hat{k} \, \phi (\vec{p} -\vec{k}/2) 
&=& S_d^3 (k^2)\, \vec{J}\cdot \hat{k} \, \vec{\epsilon} \cdot
\hat{k}~.
\end{eqnarray}
In the photoproduction, one is of course only sensitive to $S_d^1$,
see e.g. ref.\cite{silas}. $\phi (p) $ denotes the 
momentum space deuteron 
wave function. It is important
to stress that $E_d$ and $L_d$ are defined with respect to the total
angular momentum $\vec{J}= \vec{L}+\vec{S}$ and not only with respect
to the deuteron spin vector $\vec{S} = (\vec{\sigma}_p + \vec{\sigma}_n)/2$. 
To properly extract the single
scattering contribution, one has to look at the transverse and
longitudinal response functions $R_{T,L}$ (the formalism is reviewed in
\cite{DT}). The form factors $S_d^i (k^2)$ are shown in fig.2 for
photon virtualities in the range from $k^2=0$ to $k^2 =-0.1$~GeV$^2$.
All of them exhibit a strong dependence on the photon virtuality, i.e.
at $k^2 \simeq -0.1\,$GeV$^2$ they suppress the single scattering
contribution by about a factor of two.

\medskip\noindent
In addition, there are two three--body contributions at
third order (see also ref.\cite{silas}). The corresponding amplitudes
read:
\begin{eqnarray}
{{\cal M}^{(tb,a)}} &=& -2 \frac{e\, g_A\, m_d \,
  M_{\pi^0}}{(2\pi)^3 \, F_\pi^3} \,
\left\langle\frac{\vec{S} \cdot \vec{\epsilon}}{{\vec q}\,'^{\,2}}
\right\rangle_{\rm wf}~,
\nonumber \\
{\cal M}^{(tb,b)} &=& 2 \frac{e\, g_A\, m_d \, M_{\pi^0}}{(2\pi)^3 \, F_\pi^3} \,
\left\langle\frac{\vec{S}\cdot(\vec{q}\,' -\vec{k})(2\vec{q}\,' -
    \vec{k})\cdot \vec{\epsilon}}
    {(({\vec q}\,' - \vec{k}\,)^2 +{M_{\pi^+}^2})
     \,\, \vec{q}\,'^2}\right\rangle_{\rm wf}~,
\end{eqnarray}
using standard kinematical notation~\cite{silas,bblmk}. Furthermore,
 $F_\pi = 92.4\,$MeV is the pion decay constant, $g_A = 1.33$ the
axial--vector coupling constant\footnote{This is the value determined
  using the Goldberger--Treiman relation with $g_{\pi N} = 13.4$. This
  value was also used in the determination of the single scattering
  amplitudes, see e.g.~ref.\cite{bkmz}. In any case, the difference
  between this value for $g_A$ and its physical value of $1.26$ is due
  to a dimension three $\pi N$ operator, which does not occur at the
  order we are working.}
and $\langle\vartheta\rangle_{\rm wf}$ indicates that 
$\vartheta$ is sandwiched
between deuteron wavefunctions.  These matrix elements have been
evaluated using a cornucopia of wavefunctions in coordinate
and momentum space. This is a non--trivial check on our calculations.
As a further check we mention that for $k^2 =0$, we recover the
corrected results of ref.\cite{silas}.
If not stated otherwise, all numbers quoted have been obtained using
the Bonn potential. It is important
to stress that the  three--body corrections turn out to
be quite independent of the wavefunction used. This implies that
the chiral perturbation theory approach, which relies on the dominance
of the pion--exchange, is useful in this context.

\medskip\noindent
We now discuss the results for the multipoles and the S--wave cross
section. First, we consider the ${\cal O}(q^3)$ calculation, being aware of the
fact that in particular the proton electric dipole amplitude at the
photon point is not well described to that order. However, the
$k^2$--dependence of the elementary amplitudes on the proton and the
neutron is largely given  by the third order
contribution~ (see the results discussed in~\cite{bkme}). 
The ${\cal O}(q^3)$ results for
$E_d$ and $L_d$ are shown in fig.~3. 
In both cases, the three--body
contribution is sizeable. As in the case of photoproduction,
graph~(tb,a) (cf. fig.~1)
totally dominates the electric dipole amplitude. This is different
for the longitudinal response, where the contribution from graph~(tb,b) is
still smaller than the one of graph~(tb,a) but of comparable magnitude. We
observe that $E_d$  varies more significantly with increasing
$|k^2|$ than  $L_d$. As illustrated by the dotted lines in fig.~3, which have been
obtained by a constant shift of the $\pi^0 n$ amplitudes by $\pm 1
\times 10^{-3}/M_{\pi^+}$, there is some sensitivity to the elementary
scattering off the neutron. To get a more realistic estimate for the
S--wave cross section, we have adjusted the values of ${X}_{0+}^{\pi^0 p}$
and ${X}_{0+}^{\pi^0 n}$ $(X=\{E,L\}$) at the photon point
to the values obtained from the fourth
order calculation. More precisely, the electric dipole amplitudes are
taken from ref.\cite{bkmz}, the $L_{0+}^{\pi^0 p}$ from the best fit
obtained in ref.\cite{bkme} and the $L_{0+}^{\pi^0 n}$ using the
resonance saturation estimate with $g_3 = -125.6$ and $X' = -0.23$,
as detailed in~\cite{bkme} (see ref.\cite{bkm2} for a more detailed
discussion on this point).
The so calculated S--wave cross section is collected
in table~1 for a photon polarization $\epsilon = 0.67$.
We also give a range, which is obtained by adding a constant shift
of $\pm 1 \times 10^{-3}/M_{\pi^+}$ to the elementary $\pi^0 n$
longitudinal multipole (but keeping the electric dipole
amplitude at the same value as before). This serves to illustrate the sensitivity
of the S--wave cross section to the so far unmeasured  $L_{0+}^{\pi^0
  n}$.
For comparison, we note that the S--wave cross section measured
on the proton~\cite{brink,distler} for $-0.10 \le k^2 \le
-0.04\,$GeV$^2$ lies between 0.15 and 0.45~$\mu$b. 
We remark that for $k^2 \simeq -0.1\,$GeV$^2$,
$a_{0d} \simeq a_0/2$, with $a_0$ the S--wave cross
section for neutral pion production off the proton. However,
the curvature  of $a_{0d} (k^2)$ is rather different from the
one for the proton case. This is partly due to the interference
of the proton and neutron amplitudes and partly
a kinematical effect, since for a given polarisation $\epsilon$ and
virtuality $k^2$, $\epsilon_L$ is larger for the proton than
for the deuteron.

\medskip\noindent
To summarize, we have considered neutral pion electroproduction off the
deuteron at threshold using Weinberg's hybrid approach of nuclear effective
field theory (i.e. using  Weinberg's chiral counting for the interaction
kernel and applying external realistic wave functions, taken here from the
Bonn potential). To third order in the chiral expansion,
we have worked out the S--wave multipoles as a function of the photon
virtuality and discussed the sensitivity of these
to the elementary $\pi^0 n$ amplitude.
This calculation is free of undetermined parameters. 
We have also calculated the S--wave cross section
and compared its behaviour to the case of the proton.
To get a better handle
on the single scattering amplitudes, we have scaled the elementary
amplitudes so that they reproduce
the fairly precise fourth order result (at the photon point). 
The trends obtained in this
short note should be substantiated by a full scale fourth order
calculation, going also above threshold.
In addition, with the precise deuteron wave functions obtained in 
ref.\cite{egm}, it will also be possible to perform the calculation
entirely within the framework of effective field theory.
Such an investigation is under way~\cite{bkm2}. It would also 
be interesting to consider this process in the scheme of
ref.\cite{ksw}, although that approach has so far not been tested
in reactions involving on--shell pions on the external legs.


\bigskip

\section*{Tables}

\renewcommand{\arraystretch}{1.3}
\begin{table}[htb]
\begin{center}
\begin{tabular}{|c||c|c|c|c|c|}
$-k^2$ [GeV$^2$] & $0.02$  & $0.04$  & $0.06$  & $0.08$ 
                & $0.10$ \\ \hline
$a_{0d}$  [$\mu$b]       &  0.11   &  0.15  & 0.17  &  0.19 & 0.20
                \\ \hline           
$a_{0d}$ range [$\mu$b]  & 0.09 -- 0.13   &  0.12 -- 0.18    
                         & 0.14 -- 0.21   &  0.16 -- 0.23   
                         & 0.17 -- 0.24    \\ 
\end{tabular}
\end{center}
\caption{S--wave cross section for the scaled single scattering
  amplitudes as explained in the text. The range is obtained by 
  adding a constant shift of $\pm 1\times 10^{-3}/M_{\pi^+}$ to
  the elementary $\pi^0 n$ longitudinal multipole.
}
\end{table}

\newpage
\section*{Figures}

\begin{figure}[H]
   \vspace{0.5cm}
   \epsfysize=4cm
   \centerline{\epsffile{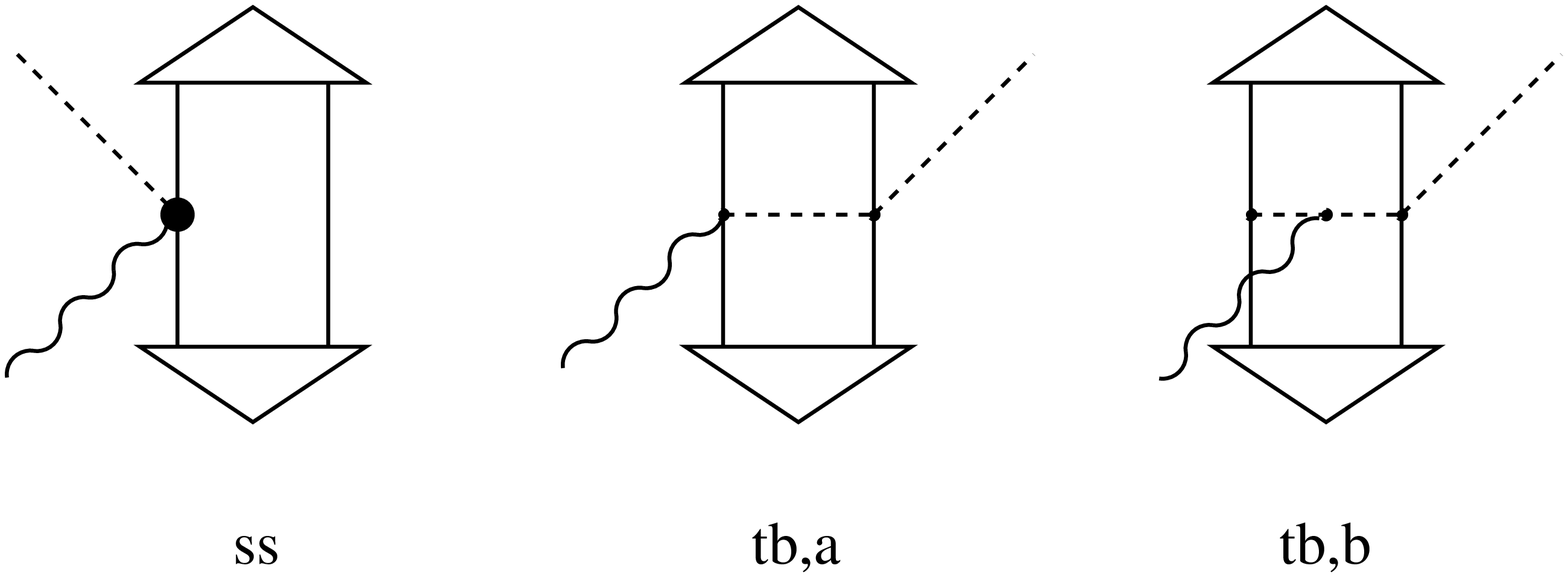}}
   \vspace{0.5cm}
   \centerline{\parbox{11cm}{\caption{\label{fig1}
Single scattering ($ss$) and three--body ($tb$) interactions 
which contribute to neutral pion electroproduction
at threshold to order $q^3$ (in the Coulomb gauge).
The solid, dashed and wiggly lines denote nucleons, pions
and photons, in order. The deuteron wavefunction is symbolized
by the triangle.
  }}}
\end{figure}

\begin{figure}[H]
   \vspace{0.5cm}
   \epsfysize=8.5cm
   \centerline{\epsffile{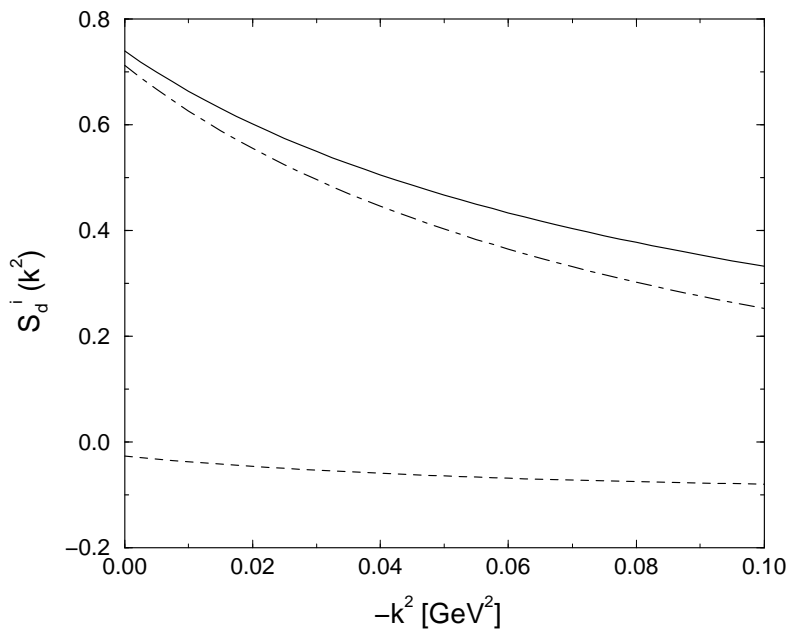}}
   \vspace{1cm}
   \centerline{\parbox{11cm}{\caption{\label{fig2}
Deuteron form factors. The solid, dashed and dot--dashed line refers
to $S_d^1 (k^2)$, $S_d^2 (k^2)$ and $S_d^3 (k^2)$, respectively.
  }}}
\end{figure}

\begin{figure}[H]
   \vspace{0.5cm}
   \epsfysize=12cm
   \centerline{\epsffile{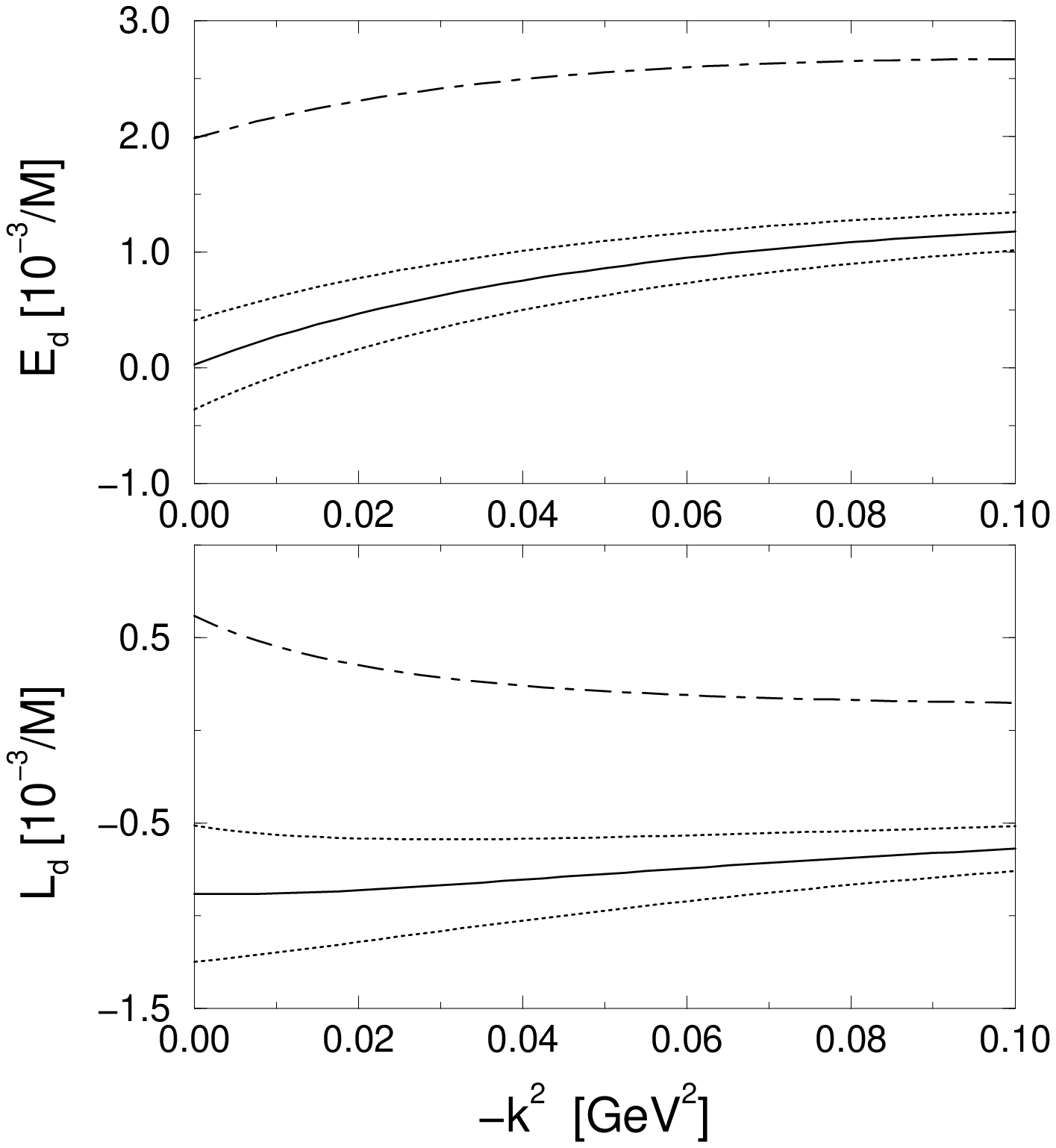}}
   \vspace{2cm}
   \centerline{\parbox{11cm}{\caption{\label{fig2a}
S--wave multipoles of the deuteron. In the upper and lower panel,
the electric dipole and the longitudinal (scalar) multipoles are
shown, respectively, by the solid lines. The dot--dashed line
is the single scattering contribution. The dotted lines are obtained
by changing the elementary neutron amplitude by $\pm 1 \times 10^{-3}
/M_{\pi^+}$ (for all values of $k^2$).
   }}}
\end{figure}


\newpage

\begin{thebibliography}{99}
\bibitem{bkmz}  V.~Bernard, N.~Kaiser and Ulf-G.~Mei{\ss}ner,
                Z. Phys. C70, 483 (1996);
                 Phys. Lett. B378, 337 (1996). 
\bibitem{bkme}  V.~Bernard, N.~Kaiser and Ulf-G.~Mei{\ss}ner,
                Nucl. Phys. A607, 379 (1996); (E) A633, 695 (1998).
\bibitem{bblmk} S.R.~Beane, V.~Bernard, T.-S.H.~Lee,
                Ulf-G.~Mei{\ss}ner and U.~van~Kolck, Nucl. Phys. A618,
                381 (1997).
\bibitem{SAL}J.C.~Bergstrom et al., Phys. Rev. C57, 3202 (1998).
\bibitem{merkel} MAMI proposal A1/1-96, A.M.~Bernstein, H.~Merkel
   et al.,  ``Threshold measurement of
  the coherent $d(e,e' d)\pi^0$ reaction''.
\bibitem{wein1} S.~Weinberg, Phys. Lett. B295, 114 (1992).
\bibitem{ksw}D.B.~Kaplan, M.J.~Savage and M.B.~Wise, Nucl. Phys. B534, 329 (1998).
\bibitem{egm} E.~Epelbaum, W.~Gl\"ockle and Ulf-G. Mei{\ss}ner,
{\tt nucl-th/9910064}, accepted for publication in Nucl. Phys. A.
\bibitem{aren} H.~Arenhoevel, Few Body Syst. 27, 141 (1999).
\bibitem{silas} S.R. Beane, C.Y. Lee and U. van Kolck, Phys. Rev. C52,
                2914 (1995).
\bibitem{DT}D.~Drechsel and L.~Tiator, J. Phys. G18, 449 (1992). 
\bibitem{bkm2}V.~Bernard, H.~Krebs and Ulf-G.~Mei{\ss}ner, forthcoming.
\bibitem{brink}H.B.~van~den~Brink et al., Nucl. Phys. A612, 391 (1997).
\bibitem{distler}M.O.~Distler et al., Phys. Rev. Lett. 80, 2294 (1998).
\end{thebibliography}
\end{document}